\documentclass[3p,twocolumn]{elsarticle}
\biboptions{sort&compress}
\usepackage{graphicx}
\usepackage{times}
\usepackage{amsmath}
\usepackage{array}
\usepackage{bm}

\begin{document}

\title{Precise determination of lattice phase shifts and mixing angles}

\author[a1]{Bing-Nan Lu\corref{cor1}}
\ead{b.lu@fz-juelich.de}
\cortext[cor1]{Corresponding author}

\address[a1]{Institute~for~Advanced~Simulation, Institut~f\"{u}r~Kernphysik, and
J\"{u}lich~Center~for~Hadron~Physics,~Forschungszentrum~J\"{u}lich,
D-52425~J\"{u}lich, Germany}

\author[a1]{Timo A. L\"ahde}

\author[a2]{Dean~Lee}
\address[a2]{Department~of~Physics, North~Carolina~State~University, Raleigh, 
NC~27695, USA}

\author[a3,a1,a4]{Ulf-G.~Mei{\ss }ner}

\address[a3]{Helmholtz-Institut f\"ur Strahlen- und
             Kernphysik and Bethe Center for Theoretical Physics, \\
             Universit\"at Bonn,  D-53115 Bonn, Germany}
\address[a4]{JARA~-~High~Performance~Computing, Forschungszentrum~J\"{u}lich, 
D-52425 J\"{u}lich,~Germany}

\begin{abstract}
We introduce a general and accurate method for determining lattice phase shifts and mixing angles,
which is applicable to arbitrary, non-cubic lattices. Our method combines angular momentum projection,
spherical wall boundaries and an adjustable auxiliary potential. This allows us to construct radial lattice
wave functions and to determine phase shifts at arbitrary energies. For coupled partial waves, we use
a complex-valued auxiliary potential that breaks time-reversal invariance.
We benchmark our method using a system of two spin-1/2 particles interacting through a finite-range potential
with a strong tensor component. We are able to extract phase shifts and mixing angles for all angular momenta
and energies, with precision greater than that of extant methods.
We discuss a wide range of applications from nuclear lattice simulations to optical lattice experiments.
%We discuss a wide range of applications from \textit{ab initio} nuclear lattice simulations to optical lattice experiments.
%that seek to emulate condensed matter systems and quantum field theories.
\end{abstract}

\begin{keyword}
%Nuclear structure, chiral effective field theory, lattice Monte Carlo
Nuclear structure, scattering phase shifts, angular momentum projection
\PACS 12.38.Gc \sep 03.65.Ge \sep 21.10.Dr
%\PACS 21.10.Dr \sep 21.30.-x \sep 21.60.De
\end{keyword}

\maketitle

\section{Introduction}
%\textit{Introduction.}$-$ 
Lattice methods are widely used in studies of quantum few- and many-body problems in
nuclear, hadronic, and condensed matter systems, see 
{\it e.g.}~Refs.~\cite{Lee2009_PPNP63-117,Drut2013_JPG40-043101,Chen2004_PRL92-257002,Borasoy2007_EPJA31-105,Borasoy2007_EPJA34-185}.
A necessary step in such studies is the computation of scattering phase shifts and mixing angles from an underlying microscopic lattice Hamiltonian.
Remarkably, the same problem arises in the context of experiments on optical lattices. Several groups have pioneered the use of ultracold atoms in optical lattices produced by
standing laser waves, to emulate the properties of condensed matter systems and quantum field theories~\cite{Hague:2011a,Hague:2015a,Li:2015a,Banerjee:2012a,Kuno:2014a}.
The basic concept is to tune the interactions of the atoms, both with each other and with the optical lattice, to reproduce the single-particle properties and particle-particle interactions
of the ``target theory''. Such studies often require a more general setup than a simple cubic lattice, for instance in the case of the hexagonal Hubbard model~\cite{Meng:2010},
which closely resembles the physics of graphene~\cite{Buividovich:2012nx} and carbon nanotubes~\cite{Luu:2015gpl}.
Clearly, a robust and accurate method for
computing scattering parameters on arbitrary lattices is needed.
%Here, we present a general solution to this problem.

For the scattering of particles on a cubic lattice,
L\"uscher's finite-volume method~\cite{Luescher1986_CMP105-153} uses periodic boundary conditions to infer elastic scattering 
phase shifts from energy eigenvalues. The method has been widely used in lattice QCD simulations  with applications to different angular
momenta~\cite{Bernard2008_JHEP08-024,Luu2011_PRD83-114508,Gockeler:2012yj,Li2013_PRD87-014502,Briceno2013_PRD88-034502} as 
well as partial-wave mixing~\cite{Briceno:2013bda}, see Ref.~\cite{Drut2013_JPG40-043101} for a recent review.  
An important advantage of L\"uscher's method is that periodic boundary conditions are typically already used in 
lattice calculations of nuclear, hadronic, ultracold atomic, and condensed matter
systems. Since no additional boundary constraints are needed, the method is easily applied to a wide class of systems.

However, L\"uscher's method requires that the finite-volume energy
levels can be accurately determined, with errors small compared 
to the separation between adjacent energy levels. This is not practical in cases such as nucleus-nucleus scattering, 
where the separation between finite-volume energy levels is many orders of magnitude smaller than the total energy of the system.
Fortunately, this problem has been solved using an alternative approach called the adiabatic projection 
method~\cite{Pine:2013zja,Elhatisari:2014lka,Rokash:2015hra,Elhatisari:2015iga}. There, initial cluster states are 
evolved using Euclidean time projection and used to calculate an effective two-cluster Hamiltonian (or transfer matrix). 
In the limit of large projection time, the spectral properties of the effective two-cluster Hamiltonian coincide with those of the original 
underlying theory. This method has been applied to nuclei and ultracold atoms, while 
applications to lattice QCD simulations of relativistic hadronic systems are currently being investigated.     

Since the adiabatic projection method reduces all scattering systems to an effective two-cluster lattice Hamiltonian, 
additional boundary conditions can be applied to the effective lattice Hamiltonian in order to compute scattering properties. 
This opens the door to methods more accurate than L\"uscher's by removing the effects of the periodic boundary conditions, 
which are otherwise a significant source of rotational symmetry breaking. One promising approach is to place the particles
in a harmonic oscillator potential and extract phase shifts from the energy eigenvalues~\cite{Luu2010_PRC82-034003,Stetcu2010_AP325-1644}.
Another prominent example is the method used in Refs.~\cite{Carlson1984_NPA424-47,Borasoy2007_EPJA34-185}, whereby a
``spherical wall'' is imposed on the relative separation between the two scattering particles. Phase shifts are
then determined using the constraint that the wave function vanish at the wall boundary. This method has
been applied to the two-nucleon problem in lattice
effective field theory (EFT)~\cite{Borasoy2008_EPJA35-343,Epelbaum2009_EPJA41-125,Epelbaum2010_EPJA45-335,Epelbaum2010_PRL104-142501}
and to lattice simulations of nucleus-nucleus scattering using the adiabatic projection method~\cite{Pine:2013zja,Elhatisari:2014lka,Rokash:2015hra,Elhatisari:2015iga}.
%and to \textit{ab initio} lattice simulations of nucleus-nucleus scattering using the adiabatic projection method~\cite{Pine:2013zja,Elhatisari:2014lka,Rokash:2015hra,Elhatisari:2015iga}.

In spite of such progress in lattice scattering theory, all methods are still lacking in precision, especially when partial-wave mixing
and high angular momenta are concerned. In previous work, numerical approximations were used for
the study of coupled-channel systems~\cite{Borasoy2007_EPJA34-185}.
We now describe an extension of the spherical wall method, which enables an efficient and precise determination of 
two-particle scattering parameters for arbitrary energies and angular momenta. We use angular momentum projection 
and solve the lattice radial equation with spherical wall boundaries, supplemented
by an ``auxiliary potential''. We test our method on a lattice model with strong tensor interactions that induce appreciable 
partial-wave mixing. We expect our method to be applicable in theoretical lattice studies of nuclear, hadronic, ultracold atomic, and condensed matter systems, 
as well as in the experimental design of optical lattices. While we discuss only non-relativistic wave mechanics in our examples here, the extension to 
relativistic systems simply entails replacing the non-relativistic dispersion relation with the relativistic one.

\section{Benchmark system}
%\textit{Benchmark system.$-$}
We begin with the eigenvalue equation
\begin{equation}
\left[-\frac{\nabla^{2}}{2\mu}+V(\bm{r},\bm{\sigma}_{1},\bm{\sigma}_{2})\right]\psi=E\psi,\label{eq:eigenvalue}
\end{equation}
where $\bm{r}$ is the relative displacement, $\bm{\sigma}_{i}$, with $i=1,2$, are the spins of the two scattering nucleons with $m_N^{} \equiv 2\mu = 938.92$~MeV.
%and $\mu$ is the reduced mass.
Following Ref.~\cite{Borasoy2007_EPJA34-185}, we take
\begin{eqnarray}
\hspace{-.5cm}
V \!\!\!\! &=& \!\!\!\! 
C\left\{ 1+\frac{r^{2}}{R_{0}^{2}}\left[3(\hat{\bm{r}}\cdot\bm{\sigma}_{1})(\hat{\bm{r}}\cdot\bm{\sigma}_{2})-\bm{\sigma}_{1}\cdot\bm{\sigma}_{2}\right]\right\} 
\nonumber \\
\!\!\!\! &\times& \!\!\!\! 
\exp\left(-\frac{r^{2}}{2R_{0}^{2}}\right),
\label{eq:interaction}
\end{eqnarray}
with $C=-2.00$~MeV and $R_{0}=2.00\times10^{-2}$~MeV$^{-1}$. We only consider
%distinguishable spin-1/2 particles and states of
states of total intrinsic spin $S=1$.
%We mainly focus on the $S=1$ channel.
%For the radial equation,
%By separating the angular and spin parts,
The radial equation is
\begin{equation}
\left[-\frac{1}{2\mu r}\frac{\partial^{2}}{\partial r^{2}}r+\frac{L(L+1)}{2\mu r^{2}}+V_{J}(r)\right]
%\left[-\frac{1}{2\mu}\frac{1}{r}\frac{\partial^{2}}{\partial r^{2}}r+\frac{L(L+1)}{2\mu r^{2}}+V_{J}(r)\right]
%%R_{J}(r)=ER_{J}(r),
\psi_{J}(r)=E\psi_{J}(r),
\label{eq:radialequation}
\end{equation}
where $L$ is the orbital angular momentum and $J$ the total angular momentum. The ``effective'' potential is
\begin{equation}
V_{J}(r)=C\left(1+\frac{2r^{2}}{R_{0}^{2}}\right)\exp\left(-\frac{r^{2}}{2R_{0}^{2}}\right),
\label{eq:uncoupledradial}
\end{equation}
for uncoupled channels, and
\begin{eqnarray}
\hspace{-.5cm}
V_{J}(r) \!\!\!\! &=& \!\!\!\!
C\left[1+\frac{r^{2}}{R_{0}^{2}}\left(\begin{array}{cc}
                        -\frac{2(J-1)}{2J+1} & \frac{6\sqrt{J(J+1)}}{2J+1}\\
                        \frac{6\sqrt{J(J+1)}}{2J+1} & -\frac{2(J+2)}{2J+1}
\end{array}\right)\right]
\nonumber \\
\!\!\!\! &\times& \!\!\!\! 
\exp\left(-\frac{r^{2}}{2R_{0}^{2}}\right),
\label{eq:coupledradial}
\end{eqnarray}
for coupled ones.
%Note that for coupled channels
%there are two independent components and the orbital angular momentum
%$L$ in
%Eq.~(\ref{eq:radialequation}) becomes a diagonal $2\times2$ matrix with elements $L=J\pm1$.
In the continuum, phase shifts and mixing angles are obtained by solving Eq.~(\ref{eq:radialequation}) using the
potentials~(\ref{eq:uncoupledradial}) and~(\ref{eq:coupledradial}) with appropriate boundary conditions.

%First, let us consider Eq.~(\ref{eq:eigenvalue}), for the case of a simple cubic lattice.
As rotational
symmetry is broken by the lattice, the energy eigenstates of Eq.~(\ref{eq:eigenvalue}) belong to the irreducible
representations (\textit{irreps}) $A_{1}$, $A_{2}$, $E$, $T_{1}$
and $T_{2}$ of the cubic group $SO(3,Z)$ rather than the
full $SO(3)$ rotational group~\cite{Borasoy2007_EPJA34-185,Johnson:1982yq,Lu2014_PRD90-034507}.
For cubic periodic boundary conditions, as in L\"uscher's method~\cite{Luescher1986_CMP105-153}, the cubic symmetry remains exact, thus our solutions can still be classified
by cubic~\textit{irreps}. Nevertheless, the rotational symmetry breaking due to the boundaries makes it difficult to identify states of high angular momentum and to
extract scattering parameters. In order to remove these effects, we
impose a hard spherical wall of radius $R_{W}$,
\begin{equation}
V\rightarrow V+\Lambda\theta(r-R_{W}),
\end{equation}
where $\theta$ is the Heaviside step function and $\Lambda$ is a (large) positive constant, intended to sufficiently suppress the wave function beyond $R_{W}$ (we set $\Lambda = 10^8$~MeV).
%We note that $\Lambda$ should be taken large enough to suppress the wave function beyond $R_{W}$ to the desired tolerance.
We take $R_W$ to exceed
the range of the interaction, such that the boundary is placed in the asymptotic (non-interacting) region. We also take $2R_W$ to be less than the difference of the box size and
the interaction range, which ensures that cubic boundary effects remain negligible.

\section{Angular momentum decomposition}
\begin{figure}
\begin{center}
\includegraphics[width=\columnwidth]{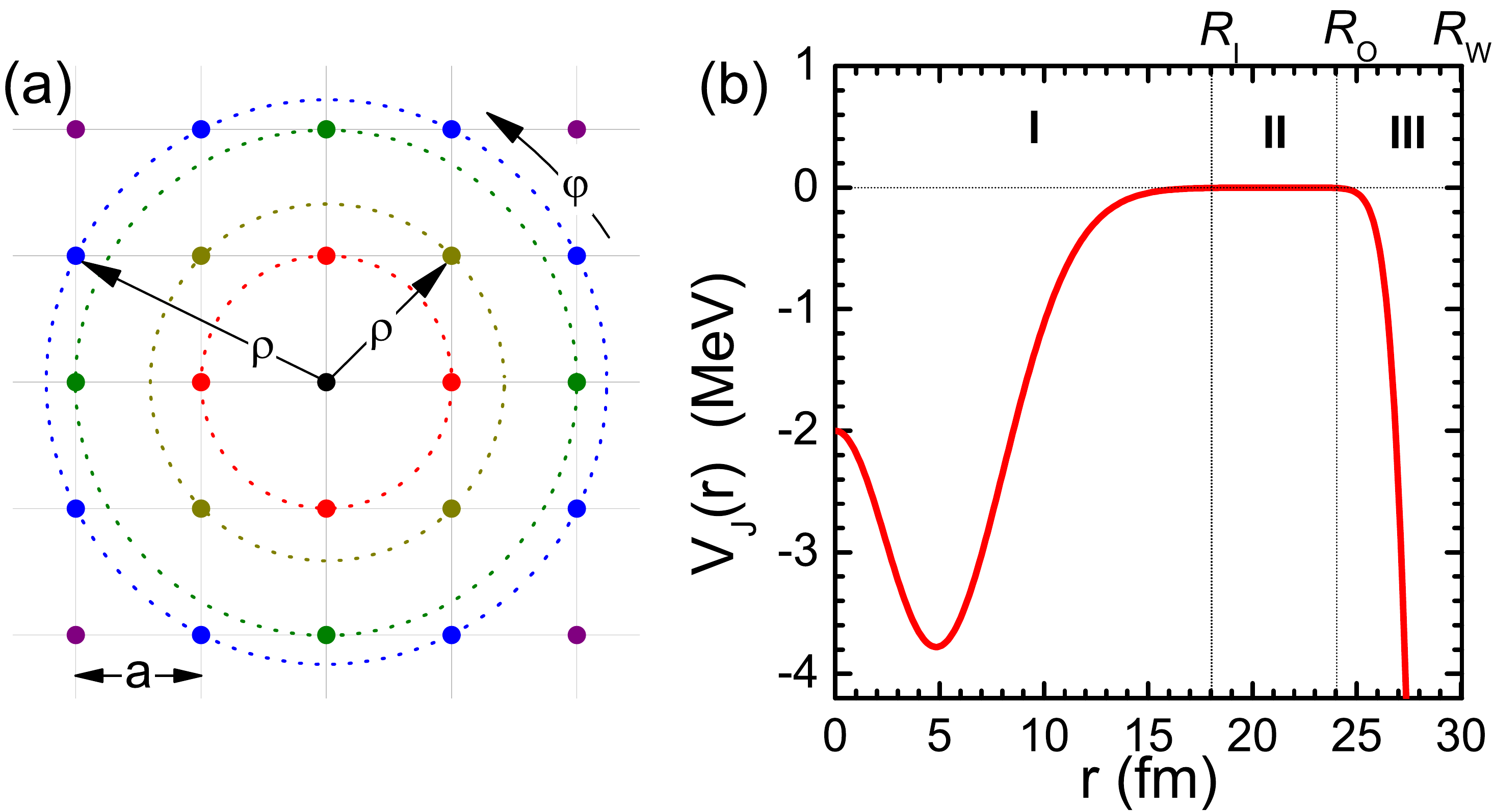}
\end{center}
%\begin{centering}
%\includegraphics[width=0.45\columnwidth]{schematic}
%\hfill
%\includegraphics[width=0.50\columnwidth]{effectivepotential}
%\includegraphics[width=0.48\columnwidth, height=0.45\columnwidth]{effectivepotential}
%\end{centering}
\caption{\label{fig:Schematic}(Color online) Left panel: Grouping of
mesh points according to lattice coordinates $(\rho,\varphi)$, with lattice spacing $a$. Right panel:
Spherical wall radius $R_W$, interaction regions I-III as discussed in the text and
effective potential $V_{J}(r)$ for uncoupled channels with $V_0=-25$~MeV.
%corresponding to Eqs.~(\ref{eq:uncoupledradial}) and~(\ref{eq:auxiliarypotential}) with $V_0=-25$~MeV.
}
\end{figure}

%\textit{Angular momentum decomposition}.$-$ 
Let $|\vec{r}\rangle
\otimes |S_z \rangle$ denote a two-body quantum state with separation $\vec r$ and $z$-component of total intrinsic spin $S_z$. We define radial lattice
coordinates $(\rho,\varphi)$ by grouping equidistant mesh points,
as shown in Fig.~\ref{fig:Schematic}.
%%%%%%%%%%%%%%%%%%%%%%%%%%%%%%%%%%%%%%%%%%%%%%%%%%%%%%%%%%%%%%%%%%%%%%%%%%%%%%%%%%%%%%%%%%%%%%
%%%%%%%%%%%%%%%%%%%%%%%%%%%%%%%%%%%%%%%%%%%%%%%%%%%%%%%%%%%%%%%%%%%%%%%%%%%%%%%%%%%%%%%%%%%%%%%
To construct radial wave functions, we
project onto states with total angular momentum $(J,J_z)$ in the continuum limit, using
\begin{eqnarray}
|m\rangle^{(J),(J_z)}_{(L)} \!\!\!\! &\equiv& \!\!\!\! 
\sum_{\vec{n},L_z,S_z}C^{J,J_z}_{L,L_z,S,S_z}Y_{L,L_z}^{\,}(\hat{n}) \nonumber \\
\!\!\!\! &\times& \!\!\!\! 
\delta_{\rho_m,|\vec{n}|}
\, |\vec{n}\rangle \otimes |S_z \rangle,
\label{eq:basis}
\end{eqnarray}
where the $Y_{L,L_z}$ are spherical harmonics with orbital angular momentum $(L,L_z)$. The $C^{J,J_z}_{L,L_z,S,S_z}$ are Clebsch-Gordan coefficients.
The parentheses around $J$, $J_z$ and $L$ on the left hand side signify that these quantum numbers are not exactly good quantum numbers.
Note that Eq.~(\ref{eq:basis}) is applicable to arbitrary geometries. Here,
$\vec n$ runs over all lattice points and the ``radial shell'' is given by the integer $m$. Then, $\rho_m$ is the distance from the origin in units of the lattice spacing $a$, and $\delta_{\rho_m,|\vec n |}$
picks out all lattice points for which $\rho_m = |\vec{n}|$.
It may be practical (especially for non-cubic lattices) to relax this condition to include all lattice points with $|\rho_m - |\vec{n}|| < \delta$ for small, positive $\delta$.
On the lattice, the $|m\rangle^{J,J_z}_L$ form a complete (but non-orthonormal) basis. We therefore compute the
norm matrix of these states before solving for the eigenstates of the lattice Hamiltonian.

We find that rotational symmetry breaking is almost entirely
due to the non-zero lattice spacing $a$.
%The spherical wall produces a negligible contribution whenever $R_W \gg a$.
As we take
%the continuum limit
$a \to 0$ at fixed $R_W$, rotational symmetry is exactly restored. The degree of mixing between different total angular
momenta $J$ and $J'$ is a useful indicator of rotational symmetry breaking. Such effects can be interpreted as arising
from the non-orthogonality of wave functions in different partial waves when their inner product is computed as a sum over discrete lattice points.
The degree of mixing is difficult to estimate {\it a priori}, as it depends strongly on the details of the interaction.

\begin{table*}
\caption{\label{tab:Energy-levels-calculated}Energy levels and differences $\Delta$ (in MeV) with
(w/) and without (w/o) unphysical $J$-mixing matrix elements. In the former case, we compute the eigenstates of the lattice Hamiltonian without
a spherical harmonic projection.}
%\centering{}%
\begin{center}
\begin{tabular}{|llccc>{\centering}p{0.3cm}llccc|}
\hline
\multicolumn{5}{|c}{Even parity} &  & \multicolumn{5}{c|}{Odd parity}\tabularnewline
\cline{1-5} \cline{7-11}
state & \hspace{-.15cm}\textit{irrep} & w/ & w/o & $\Delta$ &  & state & \hspace{-.15cm}\textit{irrep} & w/ & w/o & $\Delta$\tabularnewline
\hline
$1{}^{3}S(D)_{1}$ & $T_{1}$ & 0.037 & 0.038 & 0.001 &  & $1^{3}P_{1}$ & $T_{1}$ & 0.917 & 0.918 & 0.001\tabularnewline
$1^{3}D_{2}$ & $E$ & 2.764 & 2.766 & 0.002 &  & $1^{3}P(F)_{2}$ & $E$ & 1.795 & 1.796 & 0.001\tabularnewline
$1^{3}D(G)_{3}$ & $T_{1}$ & 3.347 & 3.351 & 0.004 &  & $1^{3}P_{0}$ & $A_{1}$ & 3.048 & 3.053 & 0.005\tabularnewline
$1^{3}G_{4}$ & $A_{1}$ & 6.562 & 6.567 & 0.005 &  & $1^{3}F_{3}$ & $A_{2}$ & 4.616 & 4.620 & 0.004\tabularnewline
$1^{3}G_{4}$ & $T_{1}$ & 6.624 & 6.637 & 0.013 &  & $1^{3}F(H)_{4}$ & $A_{1}$ & 4.998 & 5.003 & 0.005\tabularnewline
\hline
\end{tabular}
\end{center}
\end{table*}

%%%%%%%%%%%%%%%%%%%%%%%%%%%%%%%%%%%%%%%%%%%%%%%%%%%%%%%%%%%%%%%%%%%%%%%%%%%%%%%%%%%%%%%%%%%%%%%%
\begin{figure}
%\begin{center}
  \includegraphics[width=\columnwidth]{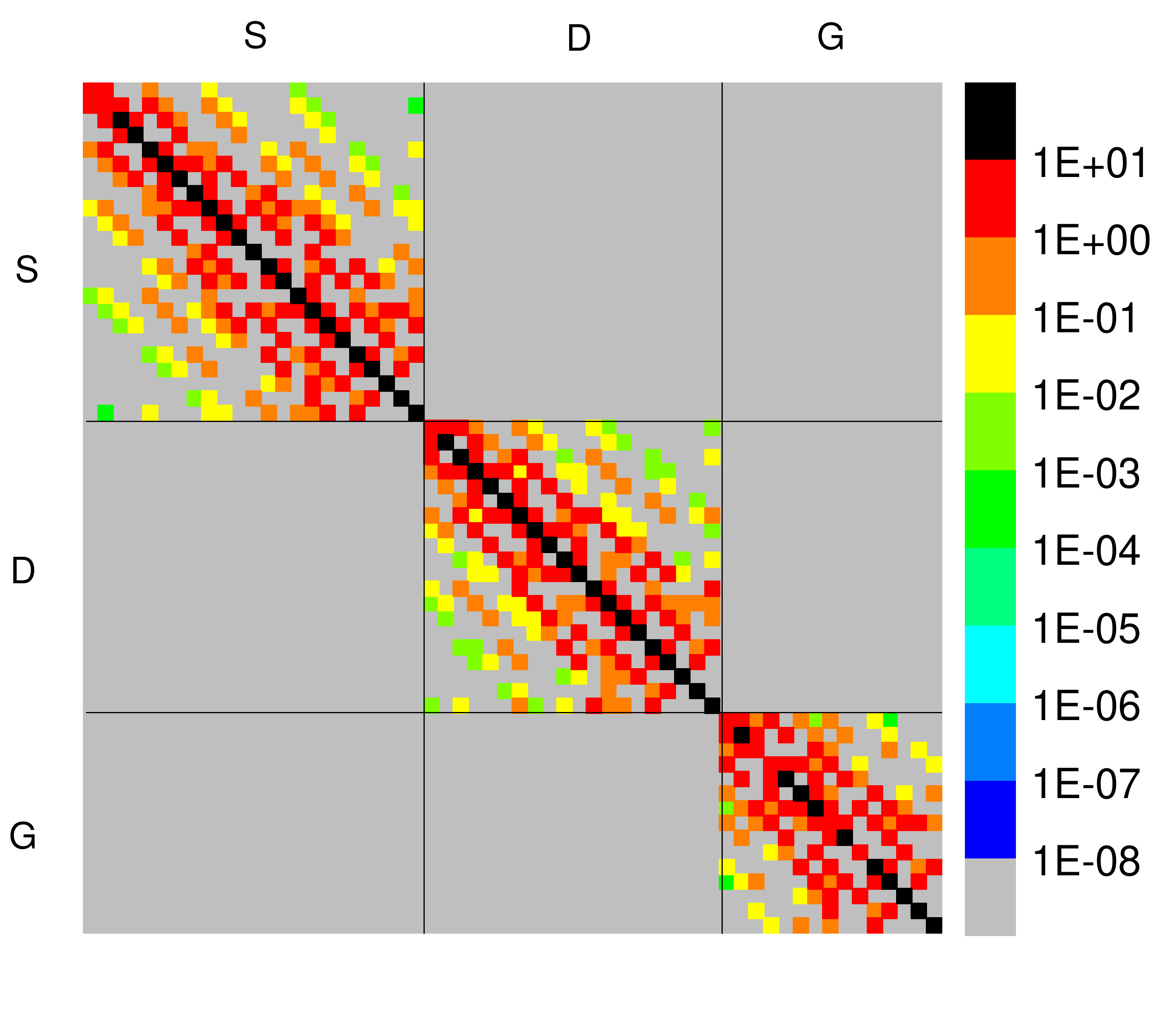}
%\end{center}
\caption{\label{fig:Hamiltonian}(Color online) Illustration of rotational symmetry breaking effects in the Hamiltonian matrix, given in the basis of Eq.~(\ref{eq:basis}). 
The colors show the magnitude of the matrix elements. To study
unphysical mixings, we remove the tensor component of $V_J(r)$. The resulting Hamiltonian matrix should ideally be block-diagonal in the $S$-, $D$- and $G$-waves {\it etc.}
Clearly, the matrix elements that cause unphysical mixings are suppressed by several orders of magnitude. 
In each block, the row and column indices represent the radial coordinates of the mesh points. For higher partial waves,
entire ``radial shells'' $\rho_m^{}$ vanish due to the angular dependence of the wave function, and such redundant rows and columns have been removed.}
\end{figure}
%%%%%%%%%%%%%%%%%%%%%%%%%%%%%%%%%%%%%%%%%%%%%%%%%%%%%%%%%%%%%%%%%%%%%%%%%%%%%%%%%%%%%%%%%%%%%%%%

Given a simple cubic lattice with a cubic-invariant interaction, unphysical $J$-mixing only occurs between cubic \textit{irreps} of the same type. If the objective is to describe a
rotationally invariant system on the lattice, then we may simply drop all unphysical couplings between channels with different $J$. We find
that rotational symmetry breaking is numerically insignificant at low energies in the spherical wall method. Still, it is instructive to study the sizes of the unphysical
$J$-mixings.  For this purpose, we use a simple cubic lattice with $a=100$~MeV$^{-1}$ and $R_{W}=10.02a$.
In the radial basis~(\ref{eq:basis}), the Hamiltonian matrix becomes nearly block-diagonal, with each block corresponding to a specific $J$.
The non-block-diagonal elements induce unphysical $J$-mixing.
%One way to assess the importance of such effects is by their influence
%on the energy eigenvalues.
In Table~\ref{tab:Energy-levels-calculated}, we examine the lowest energy levels with and without
%The columns denoted ``w/'' and ``w/o'' show the results
%with and without
$J$-mixing matrix elements. When $J$-mixing is included, we solve directly for the eigenstates of the lattice Hamiltonian without
a spherical harmonic projection.
%We find that the energy differences are numerically small.
%The energy differences are denoted by ``$\Delta$'', and we find that they are numerically small.
In Fig.~\ref{fig:Hamiltonian}, we show the Hamiltonian matrix elements in the projected basis defined in Eq.~(\ref{eq:basis}). In order to focus entirely
on unphysical mixings caused by rotational symmetry breaking, we have neglected the tensor component of $V_J^{}(r)$ in Fig.~\ref{fig:Hamiltonian}.
The magnitude of such unphysical mixing matrix elements is found to be greatly suppressed.

%In order to study the effect of this projection to a single value of $J$,
%we calculate the energy eigenvalues for our benchmark interacting system
%at a lattice spacing $a=100$ MeV$^{-1}$. The spherical wall is
%set to $R_{W}=10.02a$. In Table~\ref{tab:Energy-levels-calculated}
%we show the lowest few energy levels. The columns denoted by ``w/''
%and ``w/o'' represent the results with and without these $J$-mixing
%matrix elements, respectively. Their differences are denoted by ``$\Delta$''.
%Clearly the differences are numerically very small.

\section{Auxiliary potential}
%\textit{Auxiliary potential}.$-$ 
We first consider uncoupled channels, where
$V$ vanishes beyond an ``inner'' radius $R_{I}$.
A hard wall at $R_W$ gives access to discrete energy eigenvalues only, and a very large box is needed
at low energies. To resolve these issues, we define an ``outer'' radius $R_{O}$, between $R_{I}$ and $R_{W}$, as shown in
Fig.~\ref{fig:Schematic}.
%In order to probe a wide energy range,
We also introduce a Gaussian
``auxiliary'' potential in region~III,
%between $R_{O}$ and $R_{W}$,
%
\begin{equation}
        V_{{\rm aux}}(r) \equiv
       V_0 \: \exp\left[-(r-R_{W})^{2}/a^{2}\right], \label{eq:auxiliarypotential}
\end{equation}
with $R_{O}\leq r\leq R_{W}$,
where the separation between $R_{O}$ and $R_W$ is chosen such that $V_{{\rm aux}}$ is negligible at $R_{O}$.
Note that $V_{{\rm aux}}$ vanishes in regions~I and~II. The energy eigenvalues can now be adjusted continuously as a function of $V_0$.
In Fig.~\ref{fig:Schematic}, we show $V_{J}(r)$ for $V_0=-25$~MeV.

%%%%%%%%%%%%%%%%%%%%%%%%%%%%%%%%%%%%%%%%%%%%%%%%%%%%%%%%%%%%%%%%%%%%%%%%%%%%%%%%%%%%%%%%%%%%%%%%
%\begin{figure}
%\begin{centering}
%%\includegraphics[width=0.85\columnwidth]{effectivepotential}
%\par\end{centering}
%\centering{}\caption{\label{fig:Effectivepotandwf}(Color online)
%Effective potential $V_{J}(r)$ for uncoupled channels, corresponding to Eqs.~(\ref{eq:uncoupledradial}) and~(\ref{eq:auxiliarypotential}) with $V_0=-25$~MeV.}
%\end{figure}
%%%%%%%%%%%%%%%%%%%%%%%%%%%%%%%%%%%%%%%%%%%%%%%%%%%%%%%%%%%%%%%%%%%%%%%%%%%%%%%%%%%%%%%%%%%%%%%%

In order to extract phase shifts, we express $\psi(r)$ in region~II as
\begin{equation}
\psi(r)\cong Ah_{J}^{-}(kr)-Bh_{J}^{+}(kr), \label{eq:sphericalbesselexpan}
\end{equation}
for $R_{I}\leq r\leq R_{O}$,
where $h_{J}^{+}(kr)$ and $h_{J}^{-}(kr)$ are spherical Bessel functions, and $k=\sqrt{2\mu E}$.
The constants $A$ and $B$ can be determined {\it e.g.} by a least-squares fit in region~II.
%in the presence of $V_{{\rm aux}}$,
%we study the wave function in region II.  We express the eigenfunctions in region II as a linear combination
%of the spherical Bessel functions $h_{J}^{+}(kr)$ and $h_{J}^{-}(kr)$,
%where $k=\sqrt{2\mu E}$ is the scattering momentum. Let us denote the wave
%function by $\psi(r)$ and write
%%
%\begin{equation}
%\psi(r)\cong Ah_{J}^{-}(kr)-Bh_{J}^{+}(kr),\quad R_{I}\leq r\leq R_{O},\label{eq:sphericalbesselexpan}
%\end{equation}
%%
%where $A$ and $B$ are constants to be determined by fitting procedure in the appropriate region.
We note that
\begin{equation}
B=SA,
\label{eq:scateq}
\end{equation}
with $S\equiv\exp(2i\delta_{J})$,
from which $\delta_{J}$ can be obtained.
%where $S$ is the scattering matrix $S\equiv\exp(2i\delta_{J})$, from which the scattering phase shifts can
%be extracted.

For coupled channels, $\psi$ has two components with $L = J \pm 1$.
%with $L=J-1$ and $L=J+1$, respectively.
Given Eq.~(\ref{eq:coupledradial}), both satisfy the spherical Bessel equation in region~II, and are therefore of
the form~(\ref{eq:sphericalbesselexpan}).
%It is then convenient to introduce the column vectors
%From Eq.~(\ref{eq:coupledradial}), we know that in region II both components satisfy the spherical Bessel equation and can be written
%in the form of Eq.~(\ref{eq:sphericalbesselexpan}) separately for each value of $L$. Now, let $A$ and $B$ denote the two-component
%column vectors consisting of these coefficients,
If we denote $A\equiv(A_{J-1},A_{J+1})^{T}$ and $B\equiv(B_{J-1},B_{J+1})^{T}$, the $S$-matrix couples
channels with $L=J\pm 1$. In the Stapp parameterization~\cite{Stapp1957_PR105-302},
\begin{eqnarray}
S &\equiv& \left[\begin{array}{cc}
\exp(i\delta_{J-1}) \\
 & \exp(i\delta_{J+1})
\end{array}\right]
\nonumber \\
&\times& \left[\begin{array}{cc}
\cos(2\epsilon_{J}) & \hspace{.48cm} i\sin(2\epsilon_{J}) \\
i\sin(2\epsilon_{J}) & \hspace{.48cm} \cos(2\epsilon_{J})
\end{array}\right] 
\nonumber \\
&\times& \left[\begin{array}{cc}
\exp(i\delta_{J-1})\\
 & \exp(i\delta_{J+1})
\end{array}\right],\label{eq:Stapper}
\end{eqnarray}
where $\epsilon_{J}$ is the mixing angle.

When solving $S$ from Eq.~(\ref{eq:scateq}) as in the uncoupled case, we encounter a subtle
problem. For a simple hard wall boundary, only one independent solution per lattice energy eigenvalue is obtained. In order to determine $S$
unambiguously, two linearly independent vectors $A$ and $B$ are needed.
In Ref.~\cite{Borasoy2007_EPJA34-185}, this
problem was circumvented by taking two eigenfunctions with approximately
the same energy and neglecting their energy difference. However, such a procedure introduces
significant uncertainties.

%%%%%%%%%%%%%%%%%%%%%%%%%%%%%%%%%%%%%%%%%%%%%%%%%%%%%%%%%%%%%%%%%%%%%%%%%%%%%%%%%%%%%%%%%%%%%%%%%
\begin{figure*}
\begin{center}
\includegraphics[width=\textwidth]{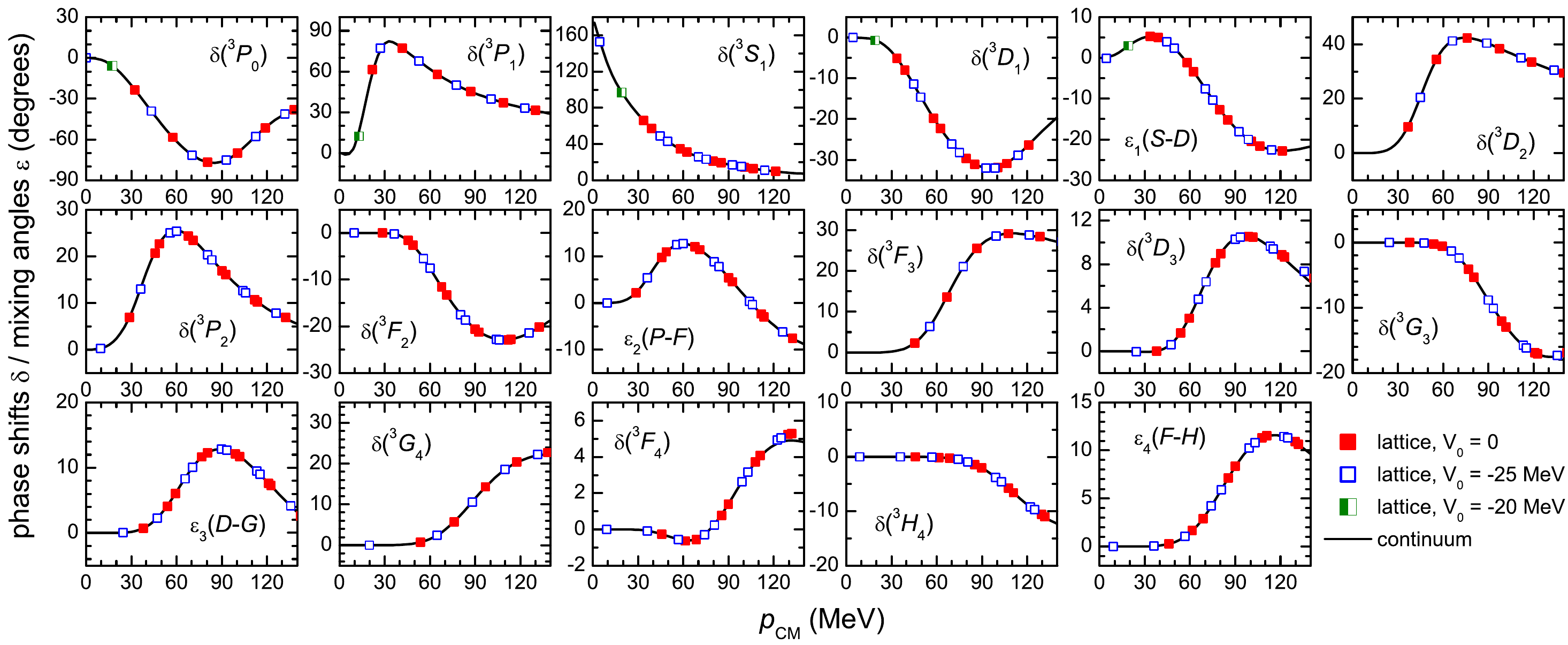}
\end{center}
\caption{\label{fig:summaryphaseshifts} (Color online)
Phase shifts and mixing angles for $J\leq4$ and $S=1$. Full, open
and ``half-open'' squares correspond to $V_0=0$, $V_0=-25$~MeV and $V_0=-20$~MeV, respectively.
For $V_0=-20$~MeV, only partial results are shown in order to reduce clutter.
Solid lines denote continuum results.}
\end{figure*}
%%%%%%%%%%%%%%%%%%%%%%%%%%%%%%%%%%%%%%%%%%%%%%%%%%%%%%%%%%%%%%%%%%%%%%%%%%%%%%%%%%%%%%%%%%%%%%%%%

%We now solve the aforementioned problem using a complex auxiliary potential.
As the potential~(\ref{eq:coupledradial}) is real
and Hermitian, an exact time-reversal symmetry results. We now add to $V_{J}(r)$ an
imaginary component,
\begin{equation}
        V_{J}(r)\rightarrow V_{J}(r)+\left[\begin{array}{cc}
                & iU_{\rm aux}(r)\\
                -iU_{\rm aux}(r)
\end{array}\right],
\end{equation}
where $U_{\rm aux}(r)$ is an arbitrary, real-valued function with
support in region~III only. This leaves $V_{J}(r)$ Hermitian and the energy eigenvalues real, while
the time-reversal symmetry is broken. Also, $\psi$
and $\psi^{*}$ are now linearly independent and satisfy Eq.~(\ref{eq:radialequation}) in regions~I and~II with identical
energy eigenvalues. In addition to Eq.~(\ref{eq:scateq}),
we have the conjugate expression,
\begin{equation}
A^{*}=SB^{*},\label{eq:stateq2}
\end{equation}
and the $S$-matrix
\begin{equation}
S=\left[\begin{array}{cc}
B & A^{*}\end{array}\right]\left[\begin{array}{cc}
A & B^{*}\end{array}\right]^{-1},\label{eq:coupledSmatrix}
\end{equation}
from~(\ref{eq:scateq}) and~(\ref{eq:stateq2}).
Phase shifts and mixing angles can then be obtained from Eq.~(\ref{eq:Stapper}).
Note that the inverse in Eq.~(\ref{eq:coupledSmatrix}) cannot be computed without $U_{\rm aux}(r)$,
since in that case $A=-B^{*}$.
%The choice of $U_{\rm aux}(r)$
%is arbitrary, and
We use
\begin{equation}
        U_{\rm aux}(r) = U_0 \, \delta_{r, r_0}, \label{eq:deltaauxiliarypotential}
\end{equation}
for $R_{O}\leq r_{0}\leq R_{W}$,
where $r_{0}$ is a radial mesh point in region~III and $U_0$ is an
arbitrary real constant. We find that the distortion of the energy eigenvalues and radial wave function introduced by this choice
is minimal.  The same methodology
we have applied here for coupled partial waves can also be applied to more general problems with different scattering constituents.

\section{Numerical results}
%\textit{Numerical results}.$-$
We benchmark our method numerically with
the interaction~(\ref{eq:interaction}) using a cubic lattice with $a~$=~100 MeV$^{-1}$ ($\pi/a=314$~MeV), box size $35a$, and
we take $R_{I} = 9.02a$, $R_{O} = 12.02a$, and $R_{W} = 15.02a$.  
%We set $R_{I}$, $R_{O}$ and $R_{W}$ to 9.02$a$, 12.02$a$ and 15.02$a$, respectively.
%The box size is $35a$, and we only discuss the $S=1$ channels.
For all
channels, we use the real auxiliary potential~(\ref{eq:auxiliarypotential}), while for coupled
channels we add the complex auxiliary potential~(\ref{eq:deltaauxiliarypotential}) with $U_0=20.0$~MeV and
$r_{0} \simeq R_{W}$.

In Fig.~\ref{fig:summaryphaseshifts}, we show our lattice phase
shifts and mixing angles. We compare with continuum
results, obtained by solving the Lippmann-Schwinger equation
for each channel.
%The results for $V_0=0$
%and $V_0=-25$ MeV are denoted by full and open squares, respectively.
%Some selected results for $V_0=-20$ MeV are denoted by ``half-open'' squares.
All our lattice results agree well with the continuum ones,
from threshold to a relative center-of-mass momentum of $p_{{\rm CM}} \equiv k =140$~MeV.
We note the marked improvement over Ref.~\cite{Borasoy2007_EPJA34-185} for the same benchmark system.

\section{Application to arbitrary lattices} 
\begin{figure}
\begin{center}
\includegraphics[width=\columnwidth]{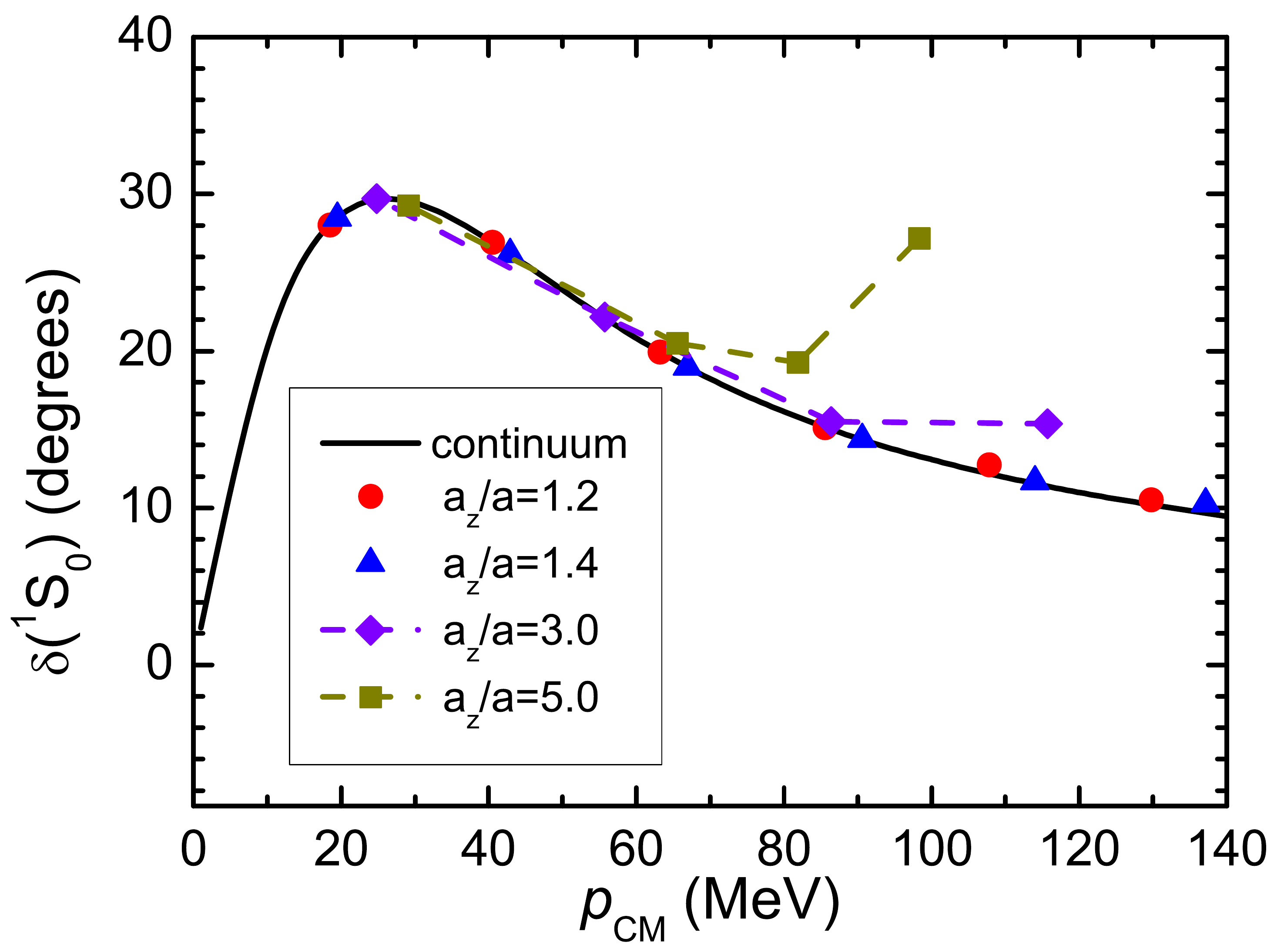}
\end{center}
\caption{\label{fig:anisotropiclattice} (Color online)
        Phase shift for the $^1S_0$ channel on anisotropic rectangular lattices.
        Circles, triangles, diamonds and squares denote results for lattice spacings $a_z=1.2a$, $a_z=1.4a$, $a_z=3.0a$ and $a_z=5.0a$, respectively.
        The dashed lines are intended as a guide to the eye.}
\end{figure}

%\textit{Application to arbitrary lattices}.$-$ 
While L{\"u}scher's method has been extended
to asymmetric rectangular boxes~\cite{Li:2007ey}, no standard method yet exists for an arbitrary lattice.
Our method can be used to characterize particle-particle interactions on arbitrary lattices, in any number of spatial dimensions.
This is significant for optical lattices, as the lattice geometry is then engineered to reproduce the single-particle energies of a given condensed matter or
quantum field theoretical system. Anisotropic lattices exhibit more breaking of rotational
invariance than a simple cubic lattice does. This is often an essential feature, {\it e.g.}\ in the crossover from a three-dimensional
system to a layered two-dimensional one. In Fig.~\ref{fig:anisotropiclattice}, we show the $^1S_0$ phase shift on an anisotropic rectangular lattice,
where the spacing along the $z$ axis, $a_z$, exceeds those along the $x$ and $y$ axes, denoted collectively by $a$.
%denoted $a_x$ and $a_y$, respectively.
The unit cell volume is $100^3$~MeV$^{-3}$ in all cases.
While we find good agreement with the continuum up to $a_z \simeq 1.4a$, this breaks down when
$a_z$ becomes comparable to the range of the interaction, with increasing deviation at high $p_{{\rm CM}}$.
%and mixing between $S$- and $D$-waves.
Such a crossover to two-dimensional behavior can be characterized in terms of mixing between the $^1S_0$ and
$^1D_2$ ($J_z = 0$) partial waves, an effect of rotational symmetry breaking. The low-energy particle-particle interactions of any lattice system can be similarly described.

%%%%%%%%%%%%%%%%%%%%%%%%%%%%%%%%%%%%%%%%%%%%%%%%%%%%%%%%%%%%%%%%%%%%%%%%%%%%%%%%%%%%%%%%%%%%%%%%%
%%%%%%%%%%%%%%%%%%%%%%%%%%%%%%%%%%%%%%%%%%%%%%%%%%%%%%%%%%%%%%%%%%%%%%%%%%%%%%%%%%%%%%%%%%%%%%%%%

\section{Summary and discussion}
%\textit{Summary and discussion}.$-$
We have described a general and systematic method for the calculation of scattering parameters on arbitrary lattices, which we
have benchmarked using a lattice model of a finite-range interaction with a strong tensor component. Extensions to more general interactions are straightforward.
The Coulomb interaction can be accounted for by replacing the spherical Bessel functions by Coulomb functions, and by defining the
distance between particles as the minimum distance on a periodic lattice. The spherical wall then removes unphysical boundary effects.  When combined with the adiabatic projection method, the techniques we have discussed can be applied to any scattering system in nuclear, hadronic, ultracold atomic, or condensed matter physics. We expect our method to be applicable to optical lattice experiments, in addition to its immediate usefulness for lattice studies
%of few- and many-body systems
in nuclear, hadronic, and condensed matter theory. In fact, the method proposed here has already been
used to significantly improve the adibatic projection method, as detailed in Ref.~\cite{Elhatisari:2016hby}.

\section*{Acknowledgments}
%\textit{Acknowledgments}.$-$ 
We are grateful for discussions with Serdar Elhatisari, Dan Moinard and Evgeny Epelbaum.
We acknowledge partial financial support from the Deutsche Forschungsgemeinschaft
(Sino-German CRC 110), the Helmholtz Association (Contract No.\ VH-VI-417),
BMBF (Grant No.\ 05P12PDTEE), the U.S. Department of Energy (DE-FG02-03ER41260),
the Chinese Academy of Sciences (CAS) President's International Fellowship Initiative (PIFI) (Grant No.\
2015VMA076) and the Magnus Ehrnrooth Foundation
of the Finnish Society of Sciences and Letters.

\section*{References}

%\bibliography{mybibfile}

\end{document}